\documentclass[preprint2,numberedappendix]{emulateapj-rtx4}
\usepackage{graphicx,graphics,amsmath}
\usepackage{bm,url}
\usepackage{color}
\def\blue{\textcolor{blue}}
\def\blue{\textcolor{black}}
\graphicspath{{./fig/}{./png/}}
\newcommand{\EQ}{\begin{equation}}
\newcommand{\EN}{\end{equation}}
\newcommand{\EQA}{\begin{eqnarray}}
\newcommand{\ENA}{\end{eqnarray}}
\newcommand{\nab}{\mbox{\boldmath $\nabla$} {}}
\newcommand{\brac}[1]{\langle #1 \rangle}

\newcommand{\DIV}{\bm{\nabla} \cdot }

\newcommand{\AAA}{\bm{{A}}}
\newcommand{\BB}{\bm{{B}}}
\newcommand{\JJ}{\bm{{J}}}
\newcommand{\UU}{\bm{{U}}}

\newcommand{\aaaa}{\mbox{\boldmath $a$} {}}

\newcommand{\meanB}{\overline{B}}

\newcommand{\meanAA}{\bm{\overline{{A}}}}
\newcommand{\meanBB}{\bm{\overline{{B}}}}
\newcommand{\meanJJ}{\bm{\overline{{J}}}}

\newcommand{\uu}{\bm{{{u}}}}
\newcommand{\bb}{\bm{{{b}}}}

\newcommand{\jj}{\bm{{{j}}}}
\newcommand{\ff}{\bm{{{f}}}}
\newcommand{\kk}{\bm{{{k}}}}
\newcommand{\xx}{\bm{{{x}}}}

\newcommand{\eee}{\hat{\mbox{\boldmath $e$}} {}}

\def\kf{k_\mathrm{f}}
\def\ii{{\rm i}}

\def\const{{\rm const}}

\def\Pm{P_\mathrm{m}}
\def\Rm{R_\mathrm{m}}

\def\Rey{\mbox{\rm Re}}

\newcommand{\mean}[1]{\overline{#1}}

\newcommand{\epsm}{\epsilon_{\rm m}}

\newcommand{\etaT}{\eta_{\rm T}}

\newcommand{\cs}{c_{\rm s}}
\newcommand{\urms}{u_{\rm rms}}

\newcommand{\Beq}{B_{\rm eq}}

\def\onethird{{\textstyle{1\over3}}}
\def\onehalf{{\textstyle{1\over2}}}

\def\onethird{{\textstyle{1\over3}}}

\newcommand{\RRRR}{\mbox{\boldmath ${\sf R}$} {}}
\newcommand{\SSSS}{\mbox{\boldmath ${\sf S}$} {}}

\newcommand{\Eq}[1]{Equation~(\ref{#1})}

\newcommand{\Sec}[1]{\S\ref{#1}}

\newcommand{\Fig}[1]{Figure~\ref{#1}}
\newcommand{\Tab}[1]{Table~\ref{#1}}
\newcommand{\Figs}[2]{Figures~\ref{#1} and \ref{#2}}

\newcommand{\bra}[1]{\langle #1\rangle}
\newcommand{\BoldVec}[1]{\mathchoice%
  {\mbox{\boldmath $\displaystyle     #1$}}%
  {\mbox{\boldmath $\textstyle        #1$}}%
  {\mbox{\boldmath $\scriptstyle      #1$}}%
  {\mbox{\boldmath $\scriptscriptstyle#1$}}%
}

\shorttitle{Hysteresis in turbulent dynamos}
\shortauthors{Karak et al.}

\begin{document}
\title{Hysteresis between distinct modes of turbulent dynamos}

\author{Bidya Binay Karak$^{1}$, Leonid L.~Kitchatinov$^{2,3}$, Axel Brandenburg$^{1,4}$}
\affil{$^1$Nordita, KTH Royal Institute of Technology and Stockholm University, Roslagstullsbacken 23, SE-10691 Stockholm, Sweden\\
$^2$Institute of Solar-Terrestrial Physics, PO Box 291, Irkutsk 664033, Russia\\
$^3$Pulkovo Astronomical Observatory, St. Petersburg, 196140, Russia\\
$^4$Department of Astronomy, Stockholm University, SE-10691 Stockholm, Sweden\\
}

\begin{abstract}
Nonlinear mean-field models of the solar dynamo show long-term variability,
which may be relevant to different states of activity inferred from
long-term radiocarbon data.
This paper is aimed to probe the dynamo hysteresis predicted by
the recent mean-field models of Kitchatinov \& Olemskoy (2010)
with direct numerical simulations.
We perform three-dimensional simulations of
large-scale dynamos in a shearing box with helically forced turbulence.
As initial condition, we either take a weak random magnetic field or
we start from a snapshot of an earlier simulation.
Two quasi-stable states are found to coexist in a certain range of parameters
close to the onset of the large-scale dynamo.
The simulations converge to one of these states depending on the
initial conditions.
When either the fractional helicity or the magnetic Prandtl number is increased
between successive runs above the critical value for onset of the dynamo,
the field strength jumps to a finite value.
However, when the fractional helicity or the magnetic Prandtl number is
then decreased again, the field strength stays at a similar value
(strong field branch) even below the original onset.
We also observe intermittent decaying phases away from the strong field
branch close to the point where large-scale dynamo action is just possible.
The dynamo hysteresis seen previously in mean-field models
is thus reproduced by 3D simulations.
Its possible relation to distinct modes of solar activity such as
grand minima is discussed.
\end{abstract}

\email{bbkarak@nordita.org}
   \maketitle

\section{Introduction}
The solar magnetic activity cycle is not a strictly periodic phenomenon.
Its duration and strength vary from cycle to cycle.
An impressive example of this aperiodicity is the famous Maunder minimum
when sunspots were extremely scarce over about 70 years \citep{HS96}.
Prolonged events of low magnetic activity like the Maunder minimum
are typical characteristics of the Sun.
Radiocarbon data reveal solar activity variations for the past $\sim11,000$ years with
27 grand minima covering about 17\% of the time
\citep{USK07,Uso13}.

Based on the extensive literature on nonlinear dynamos displaying
long-term variability, we can classify two broad theories of
grand minima: amplitude modulation through nonlinearity \citep{Spi77,Tav78,Ruz81}
and externally imposed noise \citep{C92}, as has been extensively
reviewed by \cite{Cha10}.
Amplitude modulation is found to exist in many nonlinear dynamo models.
This can result from the coupling between various dynamo modes with close
frequencies \cite[e.g.,][]{Bra89b,Bra89a,SN94,BTW98,Bro98},
and/or from the interaction between magnetic field and
differential rotation \citep{Kit94, Kuk99}.
However, the latter is less likely to apply to the Sun
on the grounds that the variation in observed differential rotation is weak.
Chaotic behavior of nonlinear dynamo models was
also identified to be a cause of amplitude modulation
in low-order models \citep[e.g.,][]{WCJ84}.
Originally this appeared to be a feature of highly truncated models,
but it was later also found in two-dimensional models \citep{CTTB98}.
On the other hand, since turbulence is the driver of dynamo action in stars,
grand minima through the resulting noise could be possible.
Indeed, dynamo coefficients such as the $\alpha$ effect
and also the Babcock--Leighton type $\alpha$ effect
\citep[through variations in the tilt angle of bipolar active regions;][]{Das10}
are known to fluctuate \citep{H88,BRRK08}.
Therefore, fluctuations in the dynamo parameters
\citep[e.g.,][]{C92,MBTT92,H93,OHS96,Cha04, Mea08,CK09,Pas14}
and even the meridional circulation \citep{Kar10} are naturally invoked
to explain the origin of grand minima.
Turbulence also introduces ``magnetic noise'' that directly affects the mean
electromotive force \citep{BS08}.
The fluctuations cause irregular changes in dynamo cycle amplitudes with
occasional wandering into states of low cycle strength.
Dynamo models with fluctuating parameters generally reproduce the grand
minima statistics \citep[e.g.,][]{CK12,KC13,OK13}.
Recent analysis of radiocarbon data by \citet{Uea14}, however, showed
that grand minima do not constitute a low-activity tail of the distribution
common for all activity cycles, but represent a separate activity mode
that cannot be interpreted as a fluctuation of the `regular' mode.
They concluded that solar dynamo regimes in grand minima and in regular
cycles are distinct.

The difference in dynamo operation between grand minima and regular
activity modes can be interpreted as a consequence of a hysteresis phenomenon
found in nonlinear mean-field dynamo models \citep{KO10}.
In the majority of such models of the solar dynamo, suppression of
poloidal field generation by the magnetic field ($\alpha$-quenching) is invoked.
This nonlinearity serves well for stabilizing magnetic field growth.
The dynamo amplification of the magnetic field takes place when the dynamo
number
\begin{equation}
    {\cal D} = \frac{\alpha\Delta\Omega R_\odot^3}{\eta_\mathrm{T}^2}
    \label{1}
\end{equation}
exceeds a critical value ${\cal D}_\mathrm{c}$,
where $\eta_{\mathrm{T}}=\eta+\eta_{\mathrm{t}}$ is the total
(microphysical plus turbulent) magnetic diffusivity,
$\Delta\Omega$ is the angular
velocity variation in the Sun, and $\alpha$ is the measure of the
$\alpha$ effect; see \cite{KR80}.
A decrease in $\alpha$ with increasing magnetic field strength
reduces the effective dynamo number to saturate the field growth.

However, not only $\alpha$ but also the eddy diffusivity $\eta_{_\mathrm{t}}$
is magnetically quenched.
Using predictions of the quasi-linear theory for $\alpha$- and
$\eta_{\mathrm{t}}$-quenching results in a non-monotonous dependence of
the effective (magnetically modified) dynamo number on the magnetic field
\citep[see Fig.\,1 in][]{KO10}.
This number initially {\em increases} with the magnetic field but the
dependence changes to a decrease for stronger fields.
If the dynamo number (\ref{1}) is increased from a subcritical
value, the saturated field amplitude jumps to a finite value of the
order of the equipartition field just after ${\cal D}$ exceeds
${\cal D}_\mathrm{c}$
and then varies smoothly with increasing ${\cal D}$ \citep{Rea94}.
If the dynamo number is then decreased, dynamo-generated finite fields
survive for ${\cal D} < {\cal D}_\mathrm{c}$, but for sufficiently small
values of ${\cal D}$, the field eventually falls to zero.
The saturated field amplitude, therefore,
depends on the pre-history of the ${\cal D}$ variation (the dynamo hysteresis).
In a finite range of ${\cal D}$-values, there are two stable solutions
with considerably different characteristic strengths of the magnetic field.
Fluctuations in dynamo parameters provoke irregular transitions between
these two solutions \citep{KO10}.
The dynamo hysteresis can thus explain the distinction between grand
minima and regular activity modes found by \citet{Uea14}.

\citet{KO10} used a mean-field dynamo model that cannot be free from
arbitrary prescriptions.
Apart from magnetic quenching of $\alpha$ and $\eta_{_\mathrm{t}}$,
there are other nonlinearities that all are implicitly present in
direct numerical simulations.
This paper probes the dynamo hysteresis with such simulations.
Using a shearing box setup, we perform simulations of helically forced turbulence,
which produce oscillating (solar-type) dynamos.
In principle, this can also be studied in realistic
global rotating magneto-convection simulations
in spherical geometry \cite[e.g.,][]{RCGBS11,Nel13,Kar15}, but those
simulations are computationally more demanding and would benefit from
guidance through simpler turbulence simulations.
By varying the amount of relative kinetic helicity, the hysteresis-type
dependence of the oscillation amplitude on the pre-history of helicity
variations is clearly seen.
Similar behavior is also found when magnetic diffusion is varied.
The simulations generally confirm the presence of two distinct regimes
of large-scale dynamos in the vicinity of the dynamo threshold.

We note that turbulent large-scale dynamos near onset have already been
studied by \citet{Rea09}, who used ABC flow forcing.
They found intermittent large-scale fields right after dynamo onset,
but in their case no cyclic dynamos were possible nor did they find
evidence of two distinct states.

\section{The Model Setup}
\label{sec:model}
In our model, we assume the fluid to be isothermal and compressible.
It obeys the equation of state $p=\cs^2 \rho$, with constant sound speed $\cs$.
Hence we solve the following equations:
\begin{equation}
\frac{D \UU}{D t} = -S U_x \hat y
-c_{\rm s}^2\nab\ln\rho
+ \rho^{-1} \left[\JJ \times \BB +
\nab\!\cdot(\!2\rho\nu\SSSS)\right]
 + \ff,
\end{equation}
\begin{equation}
\frac{D\ln\rho}{D t} =-\nab\cdot\UU,
\end{equation}
\begin{equation}
\frac{\partial \AAA}{\partial t} + \mean \UU^{(S)} \cdot \nab \AAA = - S A_y \hat x + \UU\times\BB +  \eta \nab^2 \AAA.
\end{equation}
Here
$D/Dt = \partial/\partial t + (\UU  + \mean \UU^{(S)})\cdot \bm\nab$
is the advective time derivative, $\mean \UU^{(S)}=(0,Sx,0)$ with
$S=\const$ is the imposed uniform large-scale shear flow,
$\nu$ is the constant kinematic viscosity,
$\AAA$ is the magnetic vector potential,
${\BB} = \nab\times {\AAA}$ is the magnetic field,
$\JJ = \mu_0^{-1} \nab\times\BB$ is the current density,
$\eta$ is the constant microscopic diffusivity, and
$\ff$ is a forcing function to be specified below.
The traceless rate of strain tensor $\SSSS$ is given by
${\sf S}_{ij} = \onehalf (U_{i,j}+U_{j,i}) - \onethird \delta_{ij} \DIV \bm{U},$
where the commas denote partial differentiation with respect to the
coordinate ($j$ or $i$).
The contribution of $\mean \UU^{(S)}$ to $\SSSS$ is omitted,
because it would only introduce a small contribution.

Turbulence is sustained by supplying energy to the system through a
forcing function $\BoldVec{f} = \BoldVec{f}(\xx,t)$, which is helical
and random in time ($\delta$-correlated).
It is defined as
\EQ
\ff(\xx,t)={\rm Re}\{N\ff_{\kk(t)}\exp[\ii\kk(t)\cdot\xx+\ii\phi(t)]\},
\label{ForcingFunction}
\EN
where $\xx$ is the position vector.
At each timestep the wavevector $\kk(t)$ randomly takes any value from
many possible wavevectors in a certain range around a given forcing wavenumber
$k_{\rm f}$.
The phase $-\pi<\phi(t)\le\pi$ also changes randomly at every timestep.
On dimensional grounds, we choose $N=f_0 c_{\rm s}(|\kk|c_{\rm s}/\delta t)^{1/2}$,
where $f_0$ is a non-dimensional forcing amplitude.
The transverse helical waves are produced via Fourier amplitudes \citep{Hau04}
\begin{equation}
\ff_{\kk}=\RRRR\cdot\ff_{\kk}^{\rm(nohel)}\quad\mbox{with}\quad
{\sf R}_{ij}={\delta_{ij}-\ii\sigma\epsilon_{ijk}\hat{k}_k
\over\sqrt{1+\sigma^2}},
\label{eq: forcing}
\end{equation}
where $\sigma$ is a measure of the helicity of the forcing; for positive maximum helicity,
$\sigma=1$.
The nonhelical forcing function,
$\ff_{\kk}^{\rm(nohel)}=
\left(\kk\times\eee\right)/\sqrt{\kk^2-(\kk\cdot\eee)^2}, \nonumber$
where $\eee$ is an arbitrary unit vector
not aligned with $\kk$. Note that $|\ff_{\kk}|^2=1$ and
$\ff_{\kk}\cdot(\ii\kk\times\ff_{\kk})^*=2\sigma k/(1+\sigma^2)$.

The fluid and magnetic Reynolds numbers and the magnetic Prandtl number are defined as
\begin{equation}
 \Rey=\urms/\nu\kf, \quad
\Rm=\urms/\eta\kf,\quad
\Pm=\nu/\eta, \quad
\end{equation}
where $\urms =\langle \uu^2 \rangle^{1/2}$ is the rms value of the velocity
in the statistically stationary state with
$\langle\cdot\rangle$ denoting the average over the whole domain and
$\kf$ is the mean forcing wavenumber.

The boundary conditions are shearing--periodic in the $x$ direction
and periodic in the $y$ and $z$ directions,
with dimensions
$L_x = L_y = L_z = 2 \pi$.
We always choose $S=-0.2$, $f_0=0.01$ and $\kf = 5 k_1$ or $3 k_1$, where $k_1=2\pi/L_x = 1$ is the smallest
possible wavenumber of the box.
We use non-dimensional units by setting $\cs=\rho_0=\mu_0=1$, where
$\rho_0 = \brac{\rho}$ is the volume-averaged density, which is constant
in time.
As initial conditions we take $\uu=\ln\rho=0$
and a small-scale low amplitude ($10^{-4}$)
Gaussian noise for $\AAA$.
All computations are performed using the {\sc Pencil Code}%
\footnote{\url{http://pencil-code.googlecode.com}}.
The grid resolution of all runs presented in this paper is $96\times96\times96$.

\section{Results}

We begin by focusing on the following two sets of simulations.
In one, we have simulations
for different values of $\sigma$, taking weak fields as
initial conditions (described in \Sec{sec:model}).
These simulations are performed to identify the onset of the dynamo.
In the other set, we take as initial conditions a snapshot of a previous simulation 
right after the onset of dynamo action.
The corresponding dynamo solution is oscillatory.
We perform several simulations by successively reducing
the helicity by a small value and using the resulting fields
of the previous simulation as the initial condition.
We continue this procedure until there is only a decaying solution.
In this way we identify the regime of dynamo hysteresis as the location
where, depending on the initial conditions, both non-decaying oscillatory
dynamos and decaying solutions are possible, i.e., the system becomes bistable.
Finally, we repeat the whole procedure in a different parameter regime
and explore the robustness of the results.

\begin{figure}[t]
\centering
\includegraphics[width=0.50\textwidth]{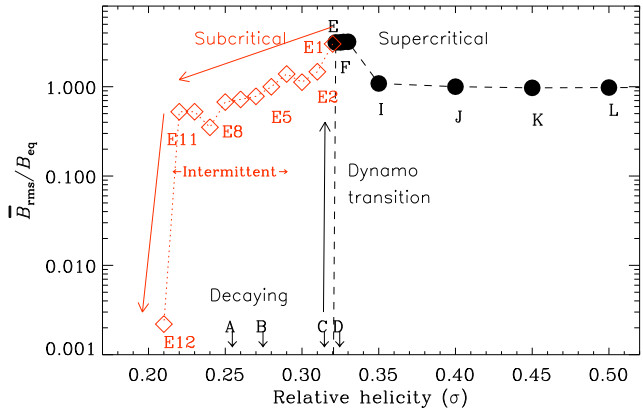}
\caption{Dynamo hysteresis, as seen in the rms value of the large-scale
magnetic field as a function of $\sigma$ (Set~I).
The filled circles (Runs~A--L) and the red diamonds (Runs~E1--E12) are from 
simulations which started with weak random seed fields 
and strong oscillatory fields of the previous simulation, respectively. 
Arrows denote the zero values for Runs~A--D.
Runs~E5--E11 show intermittent behavior.
}\label{fig:hys1}
\end{figure}
\begin{figure}[t]
\centering
\vspace{0.2in}
\includegraphics[width=0.50\textwidth]{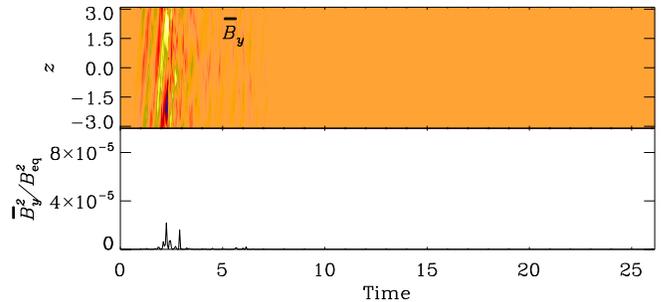}
\caption{Top: $\meanB_y$,
bottom: its time series at an arbitrarily chosen mesh point as a function
of time,
normalized by the diffusive time $(k_1^2\eta)^{-1}$.
These results are from a simulation started with weak seed field at $\sigma=0.32$,
which is just before the onset of dynamo action (Run~D).
}\label{fig:decay}
\end{figure}

\begin{table*}[t!]
\centering
\caption[]{Summary of the Runs}
      \label{tab:runs}
      \vspace{-0.5cm}
     $$
         \begin{array}{ccccccccl|ccccccccl}
           \noalign{\smallskip}
\multicolumn{9}{c|}{\rm Set~I~ (\sigma~varied)} & \multicolumn{8}{c}{\rm Set~II ~(\mathrm \Pm~varied)} \\ \hline
\rm Run &~\sigma&\urms/\cs&\Rm & D &\tilde{\meanB}&\tilde{\meanB}_{\rm x}&\tilde{\meanB}_{\rm y} &{\rm Osc} ~&~ {\rm Run} &~\Pm\;\; &\urms/\cs&\Rm&D  &\tilde{\meanB}&\tilde{\meanB}_{\rm x}&\tilde{\meanB}_{\rm y}&{\rm Osc}\\ \hline
\rm A & 0.25 & 0.32 & 12.7 &  3.28 & 0.00 & 0.00 & 0.00 & \rm N   & \rm N & 0.062 & 0.23 &  0.6 &  0.80 & 0.00 & 0.00 & 0.00 & \rm N \\
\rm B & 0.27 & 0.30 & 11.9 &  4.10 & 0.00 & 0.00 & 0.00 & \rm N   & \rm O & 0.100 & 0.34 &  1.3 &  0.86 & 0.00 & 0.00 & 0.00 & \rm N \\
\rm C & 0.31 & 0.35 & 13.6 &  3.33 & 0.00 & 0.00 & 0.00 & \rm N   & \rm P & 0.167 & 0.28 &  1.8 &  2.49 & 0.00 & 0.00 & 0.00 & \rm N \\
\rm D & 0.32 & 0.32 & 12.5 &  4.26 & 0.00 & 0.00 & 0.00 & \rm N   & \rm Q & 0.250 & 0.31 &  3.1 &  3.07 & 0.00 & 0.00 & 0.00 & \rm N \\
\mathbf {E} & \mathbf {0.322} & \mathbf {0.12} &  \mathbf {4.8} &  \mathbf {22.0} & \mathbf {3.14} & \mathbf {0.21} & \mathbf {3.08} & \mathbf {Y}   & \rm R & 0.263 & 0.27 &  2.8 &  4.35 & 0.00 & 0.00 & 0.00 & \rm N \\
\rm F & 0.325& 0.12 &  4.9 &  22.1 & 3.14 & 0.21 & 3.09 & \rm Y   & \rm S & 0.278 & 0.25 &  2.7 &  5.57 & 0.00 & 0.00 & 0.00 & \rm N \\
\rm G & 0.327& 0.12 &  4.8 &  22.7 & 3.17 & 0.21 & 3.16 & \rm Y   & {\rm \mathbf {T}} & \mathbf {0.294} & \mathbf {0.16} &  \mathbf {1.8} &  \mathbf {6.70} & \mathbf {1.68} & \mathbf {0.23} & \mathbf {1.59} & \rm \mathbf {Y} \\
\rm H & 0.33 & 0.12 &  4.8 &  23.0 & 3.19 & 0.21 & 3.14 & \rm Y   & \rm U & 0.312 & 0.18 &  2.3 &  5.21 & 1.51 & 0.20 & 1.48 & \rm Y\\
\rm I & 0.35 & 0.21 &  8.3 &  7.31 & 1.09 & 0.18 & 0.93 & \rm Y   & \rm V & 0.357 & 0.19 &  2.6 &  5.74 & 1.66 & 0.22 & 1.75 & \rm Y\\
\rm J & 0.40 & 0.22 &  8.6 &  7.47 & 1.00 & 0.20 & 0.76 & \rm Y   & \rm W & 0.500 & 0.19 &  3.8 &  7.18 & 2.03 & 0.29 & 2.93 & \rm Y\\
\rm K & 0.45 & 0.22 &  8.7 &  7.79 & 0.97 & 0.21 & 0.73 & \rm Y   & \rm X & 0.556 & 0.23 &  5.0 &  6.14 & 1.12 & 0.22 & 0.93 & \rm Y\\
\rm L & 0.50 & 0.23 &  8.9 &  7.89 & 0.98 & 0.22 & 0.71 & \rm Y   & \rm Y & 0.625 & 0.23 &  5.8 &  6.51 & 1.05 & 0.23 & 0.81 & \rm Y\\
\rm M & 1.00 & 0.23 &  9.1 &  9.14 & 0.97 & 0.27 & 0.64 & \rm Y   & \rm Z & 0.714 & 0.24 &  6.7 &  6.69 & 1.03 & 0.23 & 0.87 & \rm Y\\
\hline
\rm E1 & 0.32 & 0.13 &  5.0 &  20.0 & 3.01 & 0.20 & 2.96 & \rm Y  & \rm T1 & 0.278 & 0.13 &  1.4 &  8.52 & 1.92 & 0.29 & 1.88 & \rm Y \\
\rm E2 & 0.31 & 0.24 &  9.6 &  4.88 & 1.47 & 0.10 & 1.43 & \rm Y  & \rm T2 & 0.250 & 0.13 &  1.3 &  7.45 & 1.69 & 0.27 & 1.65 & \rm Y \\
\rm E3 & 0.30 & 0.28 & 11.2 &  3.30 & 1.13 & 0.09 & 1.06 & \rm Y  & \rm T3 & 0.200 & 0.13 &  1.0 &  6.25 & 1.34 & 0.26 & 1.29 & \rm Y \\
\rm E4 & 0.29 & 0.25 & 10.0 &  4.16 & 1.39 & 0.10 & 1.37 & \rm Y  & \rm T4 & 0.179 & 0.14 &  1.0 &  5.29 & 1.10 & 0.23 & 1.04 & \rm Y \\
\rm E5 & 0.28 & 0.30 & 11.9 &  2.74 & 1.00 & 0.08 & 0.93 & \rm Int  & \rm T5 & 0.167 & 0.15 &  1.0 &  4.54 & 0.92 & 0.20 & 0.86 & \rm Y \\
\rm E6 & 0.27 & 0.34 & 13.4 &  2.12 & 0.78 & 0.07 & 0.72 & \rm Int  & \rm T6 & 0.161 & 0.15 &  0.9 &  4.43 & 0.88 & 0.20 & 0.82 & \rm Y \\
\rm E7 & 0.26 & 0.36 & 13.9 &  1.91 & 0.72 & 0.06 & 0.66 & \rm Int  & \rm T7 & 0.156 & 0.33 &  2.0 &  1.60 & 0.00 & 0.00 & 0.00 & \rm N \\
\rm E8 & 0.25 & 0.37 & 14.7 &  1.77 & 0.67 & 0.06 & 0.63 & \rm Int  & \rm T8 & 0.154 & 0.35 &  2.1 &  1.44 & 0.00 & 0.00 & 0.00 & \rm N \\
\rm E9 & 0.24 & 0.34 & 14.6 &  1.97 & 0.35 & 0.04 & 0.29 & \rm Int  & \rm T9 & 0.152 & 0.32 &  1.9 &  1.71 & 0.00 & 0.00 & 0.00 & \rm N \\
\rm E10& 0.23 & 0.33 & 12.8 &  2.36 & 0.52 & 0.05 & 0.45 & \rm Int  & \rm T10& 0.147 & 0.21 &  1.2 &  3.25 & 0.00 & 0.00 & 0.00 & \rm N \\
\rm E11& 0.22 & 0.28 & 11.0 &  3.17 & 0.52 & 0.05 & 0.43 & \rm Int  & \rm T11& 0.143 & 0.39 &  2.2 &  1.06 & 0.00 & 0.00 & 0.00 & \rm N \\
\rm E12& 0.21 & 0.32 & 12.5 &  2.93 & 0.00 & 0.00 & 0.00 & \rm N  & \rm T12& 0.132 & 0.31 &  1.6 &  1.47 & 0.00 & 0.00 & 0.00 & \rm N \\
\hline
         \end{array}
     $$
\tablecomments{
Runs~A--E12 belong to Set~I in which
$\sigma$ is varied. Runs A--M are started from weak seed fields.
Run E1 is performed from a snapshot of Run~E (bold), but
at slightly reduced $\sigma$.
A similar procedure is continued from
Runs~E1 $\rightarrow$ E2 $\rightarrow$ E3 ... $\rightarrow$ E12.
Runs~N--T12 belong to Set~II in which $\Pm$ is varied.
Runs~N--Z are started from weak seed fields.
Run~T1 is started from a snapshot of Run~T (bold), but at
decreased $\Pm$, and a similar procedure is performed for
Runs~T1 $\rightarrow$ T2 $\rightarrow$ T3 ... $\rightarrow$ T12.
$\tilde{\meanB}=\meanB_{\rm rms}/\Beq$,
  $\tilde{\mean{B}}_x=\brac{\mean{B}_x^2}^{1/2}/\Beq$, and
  $\tilde{\mean{B}_y}=\brac{\mean{B}_y^2}^{1/2}/\Beq$.
The columns `Osc' indicate whether there are oscillations (Y) or not (N)
and `Int' denotes intermittent behavior.
}\end{table*}

\subsection{Onset of dynamo action}

\begin{figure}[t]
\centering
\vspace{0.25in}
\includegraphics[width=0.5\textwidth]{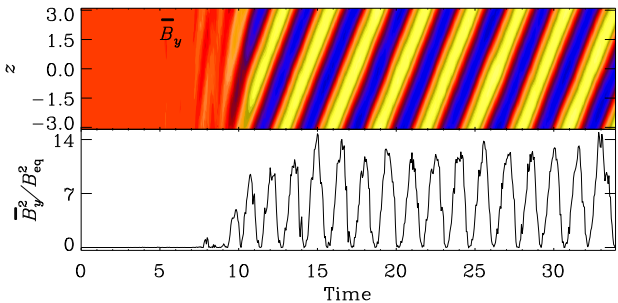}
\caption{Results from the simulation started from weak seed field at $\sigma=0.322$,
which is just right after the dynamo transition (Run~E).
The format is the same as \Fig{fig:decay}.
}\label{fig:dynamo}
\end{figure}

We perform a set of simulations by increasing the strength of the helicity
parameter $\sigma$ of the turbulent forcing, starting from 0 to 1 (Set~I).
For this set we fix $\eta = \nu = 0.005$ and $\kf = 5 k_1$.
Runs A--M in \Tab{tab:runs} show these simulations.
Along with other important parameters, we show a rough measure of the
dynamo number defined as $D = C_\alpha C_\Omega$,
where $C_\alpha=\alpha_0/\etaT k_1$, and $C_\Omega = |S|/\etaT k_1^2$,
with $\alpha_0=-\frac{1}{3}\tau \langle \boldsymbol \omega \cdot \uu \rangle$,
$\tau=(\urms\kf)^{-1}$, $\etaT=\eta + \eta_{\rm t0}$
and $\eta_{\rm t0}=\frac{1}{3}\tau \langle \uu^2 \rangle$.
In \Fig{fig:hys1} we show the temporal mean in the statistically
stationary state of the large-scale magnetic field over the whole domain,
$\mean{B}_{\rm rms}= \brac{\brac{B_x}_y^2 + \brac{B_y}_y^2+\brac{B_z}_y^2}_{xzt}^{1/2}$
normalized by $B_{\rm eq} =\urms$.
We see that the large-scale field is zero when $\sigma$ is below about 0.32, implying there
is no dynamo action.
For $\sigma=0.32$ (Run~D) in \Fig{fig:decay} we show the spatio-temporal variation of the
$y$-component of the mean magnetic field $\meanB_y=\brac{B_y}_{xy}$ (which corresponds to the toroidal
field in spherical coordinates)
and the time series of $\meanB_y^2$ at an arbitrarily chosen mesh point, normalized by $\Beq^2$ (which may be
considered as a measure of sunspot number).
Here we do not see clear magnetic oscillations.
A few cycles started to appear at around $t=2$,
but they did not survive.
The overall field is also very weak.
On increasing $\sigma$ slightly we observe a dynamo transition at $\sigma = 0.322$
(Run~E)
and the magnetic field becomes strong ($\mean{B}_{\rm rms}>\Beq$). Hence the
critical value of $\sigma$ for dynamo action is $\sigma_\mathrm{c} \approx 0.322$.
The spatio-temporal variation for this case is shown in \Fig{fig:dynamo},
where we see clear magnetic cycles with dynamo wave propagation
along the positive $z$ direction.
Together with positive helicity (which results in a negative $\alpha$) and
negative shear, migration in the positive $z$ direction is indeed expected.
However, the cycles are not regular; the amplitude varies from cycle to cycle,
similar to the observed solar cycle.
We recall that in stochastically forced mean-field dynamo models,
the cycle irregularity
is related to the amount of {\it imposed} fluctuations and the corresponding 
coherence times \citep[see examples in][]{Cha10,KC11}.
In the present simulations, however, this cycle irregularity is
naturally coming because of the finite number of eddies
and it is directly related to the scale-separation ratio ($\kf/k_1$),
which has been demonstrated in \citet{BG12}.
For $\sigma > \sigma_{\mathrm{c}}$ we always observe clear magnetic cycles and
the value of $\meanB_{\rm rms}$ remains around $\Beq$; see \Fig{fig:hys1}.

\begin{figure}[t]
\centering
\vspace{0.25in}
\includegraphics[width=0.5\textwidth]{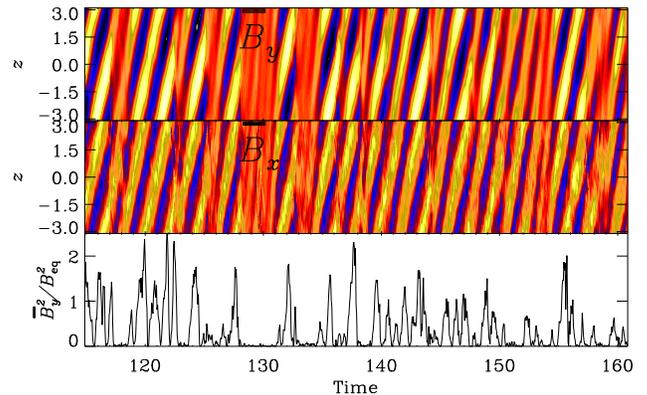}
\caption{An example of a subcritical dynamo in the bistable state of
\Fig{fig:hys1}: The simulation started from strong initial field
at $\sigma=0.22$, just before the decaying solution (Run~E11).
The format is the same as \Fig{fig:decay}, but here the $x$ component
of mean-field $\meanB_x$ is also displayed in the middle panel.
}
\label{fig:bistab1}
\end{figure}

\subsection{Dynamo hysteresis}

Now we take a snapshot of a simulation at
$\sigma =\sigma_\mathrm{c}= 0.322$ (Run~E, which is shown in
\Fig{fig:dynamo}) and perform a set of simulations by reducing $\sigma$ slowly and taking the
output of the previous simulation as the initial conditions.
Runs E1--E12 in \Tab{tab:runs} represent such cases
and the corresponding $\meanB_{\rm rms}$ are shown in \Fig{fig:hys1} as red diamonds.
We observe oscillatory solutions
over a broad parameter range, $0.22 \leq \sigma < 0.322$, in which there are otherwise
decaying solutions when started from weak fields.
Therefore, in this range,
the results depend on the initial conditions, i.e., system becomes bistable.
All the simulations are run for a sufficiently long time to ensure that they remain
in the same state.
\blue{
We recall that \cite{BCT01} studied the linear and nonlinear dynamo properties
using time-dependent ABC flows forcing in triply periodic Cartesian geometry.
Their simulations are similar to those of \cite{Rea09}, but for an
incompressible fluid.
In the nonlinear regime, \cite{BCT01} found two distinct classes of
behavior depending on the initial hydromagnetic properties of the forced
ABC flow, similar to earlier results by \cite{FRR99} in spherical geometry.
One produces the stationary solution followed by an initial exponential growth of
the magnetic field, whereas the other initially produces a dynamo solution but later 
turns into a decaying one because the flow itself evolves to a non-dynamo stage through hydrodynamic instability.
However, our study of hysteresis is different from \cite{BCT01} because we 
take different initial conditions for velocity as well as magnetic fields
and we believe that the magnetic quenching rather than the hydrodynamic instability 
is the cause of the bistability.}

In \Fig{fig:bistab1} we show the magnetic oscillations from the last run
(Run~E11) at $\sigma=0.22$, below
which the oscillations die completely.
We see that the magnetic cycles persist most of the time
in this simulation. The interesting feature is that occasionally
some of the cycles disappear or become weaker.
By comparing the first two panels of \Fig{fig:bistab1} we note that
during weaker cycles (for example, at $t \sim 130$ and $154$),
$\meanB_x$ is not reduced as much as $\meanB_y$, implying that the $\alpha$
effect dominates over the $\Omega$ effect.
This kind of intermittent behavior \blue{somewhat} resembles
the grand minima observed in the Sun.
We see a similar behavior for many runs in the bistable region,
particularly in Runs E5--E11.

\subsection{Dynamo hysteresis: dependence on $\Pm$}

To explore the robustness of the existence of dynamo hysteresis, we repeat the
same procedure in different parameter regimes of the simulations.
Here we fix $\sigma$ at 1 (fully helical flow)
but vary the magnetic diffusivity $\eta$ in each simulation;
see Runs N--Z of Set~II in \Tab{tab:runs}.
Hence, $\Pm$ varies, but $\nu=0.005$ is unchanged.
This is similar to the experiments of \citet{Rea09}, who used ABC flow forcing.

\begin{figure}[t]
\centering
\includegraphics[width=0.5\textwidth]{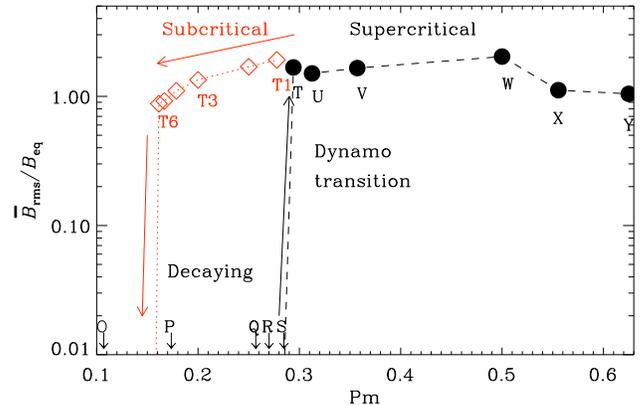}
\caption{
Similar to \Fig{fig:hys1}, but from a different set of simulations (Set~II, Runs N--T12; see also \Tab{tab:runs})
where $\eta$ is varied while $\sigma$ and $\nu$ are held fixed.
}\label{fig:hys2}
\end{figure}

The black points in \Fig{fig:hys2} show $\meanB_{\rm rms}$ from different
simulations started with weak seed fields as the initial conditions (Runs N--Z).
We see that when $\Pm$ just exceeds about 0.29, the
dynamo is excited and the magnetic field becomes oscillatory.
The critical $\Pm$ for dynamo action is $P_{\mathrm{m}}^{\mathrm{c}} \approx 0.294$.

\begin{figure}[t]
\centering
\includegraphics[width=0.5\textwidth]{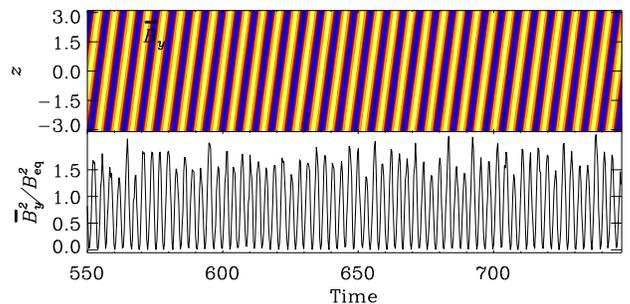}
\caption{Example of a subcritical dynamo in the bistable state of \Fig{fig:hys2}:
the simulation started from a strong initial field
at $\Pm=0.1613$, just above the value for the decaying solution (Run~T6).
The format is the same as in \Fig{fig:decay}.
}\label{fig:bist2}
\end{figure}

\begin{table}[t!]
\centering
\caption[]{Same as Set~II in \Tab{tab:runs} but simulations are performed at forcing wavenumber $\kf=3$ and $\nu=8\times10^{-3}$.}
      \label{tab:runs3}
      \vspace{-0.5cm}
     $$
         \begin{array}{p{0.05\linewidth}lcccccrrcl}
           \noalign{\smallskip}
\multicolumn{9}{c}{\rm Set~III} \\ \hline
Run        &~~\Pm &\urms/\cs&\Rm & D &\tilde{\meanB}&\tilde{\meanB}_{\rm x}&\tilde{\meanB}_{\rm y} &{\rm Osc} \\ \hline
A$^\prime$ & ~~0.50 & 0.28 &  5.5 &  2.63 & 0.00 & 0.00 & 0.00 & \rm N\\
B$^\prime$ & ~~1.00 & 0.38 & 15.0 &  1.71 & 0.00 & 0.00 & 0.00 & \rm N\\
C$^\prime$ & ~~1.14 & 0.38 & 17.4 &  1.76 & 0.00 & 0.00 & 0.00 & \rm N\\
D$^\prime$ & ~~1.33 & 0.32 & 17.2 &  2.77 & 0.00 & 0.00 & 0.00 & \rm N\\
\bf {E}$^\prime$ & ~~\mathbf {1.60} & \mathbf {0.14} &  \mathbf {9.1} & \mathbf {14.81} & \mathbf {2.23} & \mathbf {0.19} & \mathbf {2.08} & {\rm \mathbf {Y}}\\
F$^\prime$ & ~~2.00 & 0.13 & 10.8 & 18.66 & 2.50 & 0.19 & 2.39 & \rm Y\\
G$^\prime$ & ~~2.67 & 0.13 & 14.0 & 23.13 & 2.59 & 0.18 & 2.36 & \rm Y\\
H$^\prime$ & ~~4.00 & 0.13 & 21.3 & 25.01 & 2.85 & 0.15 & 2.63 & \rm Y\\
I$^\prime$ & ~~8.00 & 0.13 & 42.6 & 29.19 & 2.52 & 0.12 & 2.25 & \rm Y\\
\hline
E$^\prime1$ & ~~1.14 & 0.13 &  6.0 & 13.99 & 2.20 & 0.22 & 2.09 & \rm Y\\
E$^\prime2$ & ~~1.00 & 0.14 &  5.5 & 11.15 & 1.95 & 0.21 & 1.85 & \rm Y\\
E$^\prime3$ & ~~0.89 & 0.14 &  4.8 & 10.74 & 1.89 & 0.22 & 1.80 & \rm Y\\
E$^\prime4$ & ~~0.80 & 0.15 &  4.7 &  8.36 & 1.72 & 0.21 & 1.61 & \rm Y\\
E$^\prime5$ & ~~0.73 & 0.14 &  4.0 &  8.79 & 1.72 & 0.22 & 1.64 & \rm Y\\
E$^\prime6$ & ~~0.67 & 0.15 &  3.9 &  7.44 & 1.56 & 0.21 & 1.48 & \rm Y\\
E$^\prime7$ & ~~0.62 & 0.14 &  3.4 &  7.59 & 1.51 & 0.22 & 1.43 & \rm Y\\
E$^\prime8$ & ~~0.57 & 0.14 &  3.1 &  7.48 & 1.46 & 0.22 & 1.36 & \rm Y\\
E$^\prime9$ & ~~0.53 & 0.15 &  3.2 &  5.64 & 1.28 & 0.21 & 1.23 & \rm Y\\
E$^\prime10$& ~~0.50 & 0.36 &  7.2 &  1.36 & 0.00 & 0.00 & 0.00 & \rm N\\
\hline
         \end{array}
     $$
\tablecomments{
Runs~E$^\prime1$--E$^\prime10$ (lower part of the table) have been
restarted from Run~E$^\prime$ (bold).
}\end{table}

\begin{figure}[t]
\centering
\includegraphics[width=0.5\textwidth]{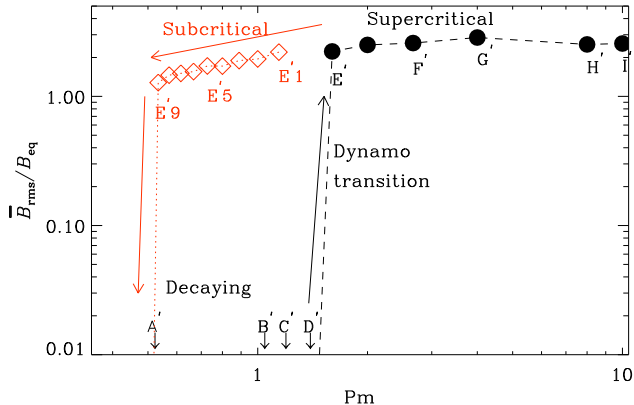}
\caption{Similar to \Fig{fig:hys2} but simulations are performed at $\kf=3$ and $\nu=8\times10^{-3}$ (Set~III in \Tab{tab:runs3}).
}\label{fig:hys3}
\end{figure}

Next, as before, we take an oscillatory dynamo solution (Run~T) as the
initial condition for the new simulation and decrease $\Pm$ by a small
value progressively in each simulation by taking as initial conditions
the last snapshot from the previous simulation.
Runs T1--T12 in \Tab{tab:runs} are such examples
and the corresponding $\meanB_{\rm rms}$ are shown as red diamonds in \Fig{fig:hys2}.
We see that, up to about $\Pm=0.16$,
we obtain an oscillatory large-scale magnetic field.
\Fig{fig:bist2} shows the typical magnetic cycles from a simulation at
$\Pm = 0.1613$ (Run~T6) below which the oscillation dies.
Note that in the bistable stage,
unlike the previous set of simulations, for example shown in \Fig{fig:bistab1}
where some of the magnetic cycles disappear occasionally, here we observe
cycles all of the time.
However when we repeat the whole procedure at $\sigma=0.5$ instead of 1, we
see this kind of intermittent behavior in the bistable regime.
To obtain even more confidence in the results we have repeated
another set of simulations (Set~III) at $\kf=3$ and $\nu=8\times10^{-3}$.
These simulations are at slightly higher $\Rm$.
\Tab{tab:runs3} gives a summary of these runs, and \Fig{fig:hys3} shows
the corresponding dynamo hysteresis.
We clearly see a similar behavior.

\section{Comparison with analytic predictions}

To understand why the hysteresis discussed in this paper has not been
seen before, we need to assess more carefully the parameter regimes
of our solutions.
Given that we use (shearing) periodic boundary conditions, there are
no magnetic helicity fluxes in or out of the domain, so we can describe
the solutions by comparing with the analytic results of
\citet[][hereafter BB02]{BB02}.
The dynamical quenching theory used in BB02 was already compared
with numerical solutions by \citet{KB09}.
One of the predictions they tested was that the saturation level
of the mean magnetic field, which they gave in the form
\begin{equation}
\frac{\meanB_{\rm rms}^2}{\Beq^2} \approx
\frac{\epsilon_{\rm f}\kf}{\epsilon_{\rm m}k_{\rm m}}
-\left(1+\frac{\eta}{\eta_{\rm t0}}\right),
\label{BfinBeq}
\end{equation}
where $k_{\rm f}$ and $k_{\rm m}$ are, respectively, the effective wavenumbers
of the fluctuating and mean fields, defined via
$k_{\rm f}^2=\bra{\jj\cdot\bb}/\bra{\aaaa\cdot\bb}$ and
$k_{\rm m}^2=\bra{\meanJJ\cdot\meanBB}/\bra{\meanAA\cdot\meanBB}$,
and $\epsilon_{\rm f}$ and $\epsilon_{\rm m}$ are their fractional
helicities, defined via
\begin{equation}
\epsilon_{\rm f}k_{\rm f}=\mu_0\bra{\jj\cdot\bb}/\bra{\bb^2},
\quad
\epsilon_{\rm m}k_{\rm m}=\mu_0\bra{\meanJJ\cdot\meanBB}/\bra{\meanBB^2}. \label{equ:epsmkm}
\end{equation}

\begin{figure}[t]
\centering
\includegraphics[width=0.5\textwidth]{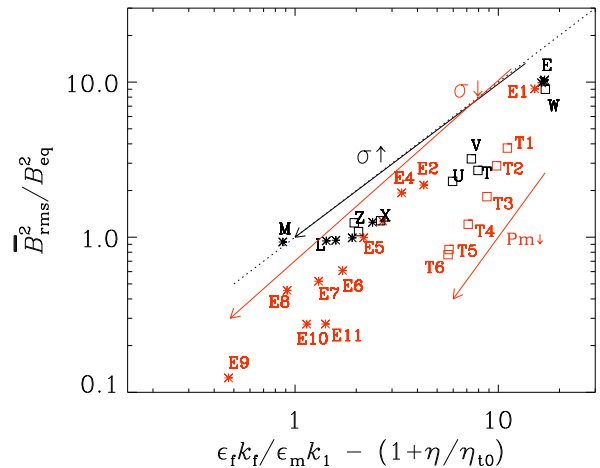}
\caption{Scatter plot between $\meanB_{\rm rms}^2/B_{\rm eq}^2$ and
$\epsilon_{\rm f}\kf/\epsilon_{\rm m}k_1 - (1+\eta/\eta_{\rm t0})$.
The dotted line shows the comparison with theory (\Eq{BfinBeq}).
The black asterisks represent data from Runs~E--M of Set~I (also represented by black points in \Fig{fig:hys1}),
whereas red asterisks are from Runs~E1--E11 (red points in \Fig{fig:hys1}).
The black squares represent data from Runs~T--Z of Set~II (also represented by black points in \Fig{fig:hys2}),
whereas red squares are from Runs~T1--T6 (red points in \Fig{fig:hys2}).
}\label{fig:epsf}
\end{figure}

\Fig{fig:epsf} shows our data from two sets of simulations (Sets~I--II) which
produce significant large-scale magnetic fields, i.e., Runs E--E11 from Set~I
and Runs T--T6 from Set II.
We have added labels to some of the data points to identify the runs.
We note that, unlike \Figs{fig:hys1}{fig:hys2}, where we have plotted
$\meanB_{\rm rms}/\Beq$, here we plot $\meanB_{\rm rms}^2/\Beq^2$, but in a smaller range,
which is why the data in \Fig{fig:epsf} show a nearly linear variation.
Even in a limited range, apart from a small offset, there is a reasonable
agreement between our data and the theory given by \Eq{BfinBeq};
see the dotted line in \Fig{fig:epsf}.
However, \citet{KB09} had data in a wider range and found better agreement
for higher field strengths.
In \Fig{fig:epsf} we observe that, when increasing the helicity
parameter $\sigma$ (from Runs~E to M in Set~I), both the wavenumber
ratio of fluctuating to mean fields, as defined by \Eq{equ:epsmkm},
and the strength of the mean field decrease.
A similar trend is followed while decreasing $\sigma$ (Runs E1--E8),
except for the last few runs (Runs~E9--E11), which deviate significantly.
Qualitatively similar behavior is observed in Set~II, when $\Pm$ is decreased
from Runs~T1 to T6, although they consistently deviate from the other runs.
However for Runs~T--Z, the trend is not monotonous.

\begin{figure}[t]
\centering
\includegraphics[width=0.5\textwidth]{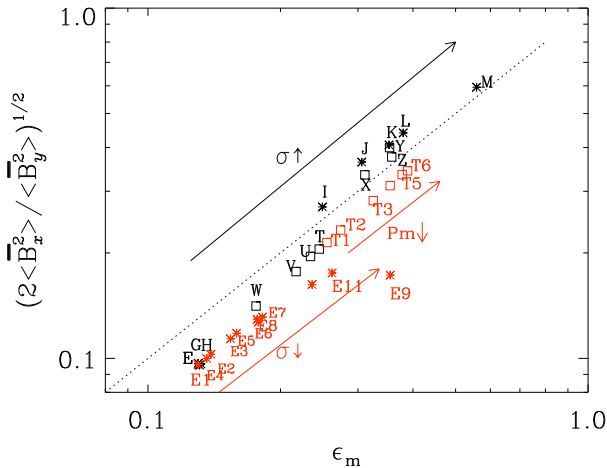}
\caption{Scatter plot between the ratio of poloidal to toroidal field and
$\epsilon_{\rm m}$.
The dotted line shows the comparison with theory (\Eq{equ:epsmkm2}).
Representation of data symbol are same as in \Fig{fig:epsf}.
}\label{fig:Bpt}
\end{figure}

Another prediction of BB02 was that $\epsm$ is directly proportional
to the ratio of poloidal to toroidal magnetic field amplitudes via
\begin{equation}
\epsilon_{\rm m}=\left(2\bra{\meanB_x^2}/\bra{\meanB_y^2}\right)^{1/2}.
\label{equ:epsmkm2}
\end{equation}
From \Eq{equ:epsmkm} we compute $\epsilon_{\rm m}$ by assuming
$k_{\rm m} = - k_{\rm 1}$ and show in \Fig{fig:Bpt} a scatter plot of
$\epsilon_{\rm m}$ versus $(2\bra{\meanB_x^2}/\bra{\meanB_y^2})^{1/2}$.
Here we see better agreement with \Eq{equ:epsmkm2}, as indicated by the dotted line.
In Set~I, with increasing $\sigma$ from Runs~E to M,
the ratio of the poloidal to toroidal field increases,
but the same happens from Runs~E1 to E11 with decreasing $\sigma$ as well,
which was unexpected.
The same trend is observed as we decrease $\Pm$
from Runs~T1 to T6 in Set~II, but in Runs~T--Z the variation is not monotonous.
Furthermore, we note that
the data points corresponding to two different regimes---subcritical and supercritical---lie
on different lines (compare red and black points in \Fig{fig:Bpt}).

We note that in \Fig{fig:epsf}, and also to some extent in \Fig{fig:Bpt},
the simulation data are systematically below the analytically expected values.
A smaller value of $\meanB_{\rm rms}$ could readily be explained as being
a combination of several modes, which results in reduced averages.
This is reminiscent of the fratricide $\alpha\Omega$ dynamos of
\cite{HRB11}, who found that they can be destroyed by their growing
$\alpha^2$ dynamo siblings.
An important difference, however, is that they never found the recovery
of the $\alpha\Omega$ dynamo.
This might be related to different orderings of the onsets
of $\alpha^2$ and $\alpha\Omega$ dynamo action, but this has
not been investigated further.

\section{Conclusion and Discussion}
As discussed in the Introduction, dynamo hysteresis predicted by the
nonlinear mean-field model of \cite{KO10} may be relevant to
the understanding of distinct modes of solar activity found by \citet{Uea14}.
Our simulations of turbulent dynamos in shearing boxes with
helically forced turbulent flows, which resemble
the equivalent $\alpha\Omega$ solar dynamo, demonstrate,
for the first time, hysteresis behavior.
By performing several simulations, either by varying the helicity parameter $\sigma$ or the
magnetic Prandtl number $\Pm$, we observe two stable states
with largely different characteristic field strengths depending
on the initial conditions of the simulation.
A decaying solution is obtained when the simulation is started with weak
random seed fields, but otherwise an oscillatory solution is obtained when
the simulation is started from a snapshot of a previous oscillatory dynamo.

We emphasize that our simulations show hysteresis close to the dynamo onset only.
This raises the question of whether the Sun may be close to the marginal state for the dynamo.
Stellar observations indicate that this might indeed be the case.
Magnetic activity is known to be correlated with rotation rate, but leads
to angular momentum loss through a magnetically coupled stellar wind
\citep{K67,HN87}.
Spin-down of solar-type stars does not continue above a certain rotation
period that depends on the spectral type \citep{R84}.
The maximum period probably corresponds to the rotation rate
where the global dynamo ceases.
The maximum period for G2 dwarfs is only slightly larger than
the rotation period of the Sun \citep[see Fig.\,1 in][]{R84}.
The stars showing low magnetic activity similar to the solar grand minima
are typically old and slow rotators \citep{SB92}.

We note that the grand minima in models based on stochastic fluctuations
\cite[e.g.,][]{H93,HST94,BS08,Mea08} always result in random
deviations from a single (regular) state.
Thanks to the reconstructed solar activity record \citep{Uea14}, however,
a different picture emerges in that the grand minima cannot be
described in terms of random fluctuations of a single solar-activity
mode, but are distinct from the regular mode and produced
as a result of sudden transitions from the regular mode to weak-field mode.

Transitions between the distinct dynamo regimes may be caused by small-scale 
hydromagnetic fluctuations inherent to 3D simulations.
The transitions do indeed happen in the simulations
with relative kinetic helicity below unity (\Fig{fig:bistab1}).

The intermittency of distinct dynamo regimes, however,
disappeared for maximally helical forcing (\Fig{fig:bist2}).
In mean-field language, an increase in fractional
helicity changes the dynamo from $\alpha\Omega$ toward $\alpha^2$ type.
There is no hysteresis for the $\alpha^2$ dynamo.
In any case, maximal helical forcing would not be realistic for the Sun.
The dependence of intermittency on the (fractional) helicity
should be explored further in future \blue{work}.

Besides the demonstration of a hysteresis phenomenon in a turbulent dynamo,
we have
analyzed the simulation data to compare with the dynamical theory of BB02.
We observe that the data from subcritical and supercritical dynamos
behave in a qualitatively similar way and that they are in reasonable
agreement with the theoretical predictions.
We end by remarking that, although our findings of hysteresis between
two distinct modes of dynamos is relevant to the recently discovered
bi-modal solar activity of \citet{Uea14}, our simulations are far from
the real Sun.
Therefore, future research is necessary to explore
similar behaviors in more realistic setups.

\begin{acknowledgements}
We thank an anonymous referee for careful review and valuable comments.
LLK is thankful to the Russian Foundation for Basic Research (project 13-02-00277) for the support
and AB acknowledges support through the Swedish Research Council grants
621-2011-5076 and 2012-5797, the Research Council of Norway under the
FRINATEK grant 231444.
The computations have been carried out at the National Supercomputer
Centres in Link\"oping and Ume{\aa}, the Center for Parallel
Computers at the Royal Institute of Technology in Sweden, and
the Nordic High Performance Computing Center in Iceland.
\end{acknowledgements}

\bibliographystyle{apj}
\bibliography{paper}

\begin{thebibliography}{53}
\expandafter\ifx\csname natexlab\endcsname\relax\def\natexlab#1{#1}\fi

\bibitem[{{Beer} {et~al.}(1998){Beer}, {Tobias}, \& {Weiss}}]{BTW98}
{Beer}, J., {Tobias}, S., \& {Weiss}, N. 1998, \solphys, 181, 237

\bibitem[{{Blackman} \& {Brandenburg}(2002)}]{BB02}
{Blackman}, E.~G., \& {Brandenburg}, A. 2002, \apj, 579, 359

\bibitem[{{Brandenburg} \& {Guerrero}(2012)}]{BG12}
{Brandenburg}, A., \& {Guerrero}, G. 2012, in IAU Symposium, Vol. 286, IAU
  Symposium, ed. C.~H. {Mandrini} \& D.~F. {Webb}, 37--48

\bibitem[{{Brandenburg} {et~al.}(1989{\natexlab{a}}){Brandenburg}, {Krause},
  {Meinel}, {Moss}, \& {Tuominen}}]{Bra89b}
{Brandenburg}, A., {Krause}, F., {Meinel}, R., {Moss}, D., \& {Tuominen}, I.
  1989{\natexlab{a}}, \aap, 213, 411

\bibitem[{{Brandenburg} {et~al.}(2008){Brandenburg}, {R{\"a}dler},
  {Rheinhardt}, \& {K{\"a}pyl{\"a}}}]{BRRK08}
{Brandenburg}, A., {R{\"a}dler}, K.-H., {Rheinhardt}, M., \& {K{\"a}pyl{\"a}},
  P.~J. 2008, \apj, 676, 740

\bibitem[{{Brandenburg} \& {Spiegel}(2008)}]{BS08}
{Brandenburg}, A., \& {Spiegel}, E.~A. 2008, Astron. Nachr., 329, 351

\bibitem[{{Brandenburg} {et~al.}(1989{\natexlab{b}}){Brandenburg}, {Tuominen},
  \& {Moss}}]{Bra89a}
{Brandenburg}, A., {Tuominen}, I., \& {Moss}, D. 1989{\natexlab{b}}, Geophys.
  Astrophys. Fluid Dynam., 49, 129

\bibitem[{{Brooke} {et~al.}(1998){Brooke}, {Pelt}, {Tavakol}, \&
  {Tworkowski}}]{Bro98}
{Brooke}, J.~M., {Pelt}, J., {Tavakol}, R., \& {Tworkowski}, A. 1998, \aap,
  332, 339

\bibitem[{{Brummell} {et~al.}(2001){Brummell}, {Cattaneo}, \& {Tobias}}]{BCT01}
{Brummell}, N.~H., {Cattaneo}, F., \& {Tobias}, S.~M. 2001, Fluid Dynam. Res.,
  28, 237

\bibitem[{{Charbonneau}(2010)}]{Cha10}
{Charbonneau}, P. 2010, Liv. Rev. Sol. Phys., 7, 3

\bibitem[{{Charbonneau} {et~al.}(2004){Charbonneau}, {Blais-Laurier}, \&
  {St-Jean}}]{Cha04}
{Charbonneau}, P., {Blais-Laurier}, G., \& {St-Jean}, C. 2004, \apjl, 616, L183

\bibitem[{{Choudhuri}(1992)}]{C92}
{Choudhuri}, A.~R. 1992, \aap, 253, 277

\bibitem[{{Choudhuri} \& {Karak}(2009)}]{CK09}
{Choudhuri}, A.~R., \& {Karak}, B.~B. 2009, Res. Astron. Astrophys., 9, 953

\bibitem[{{Choudhuri} \& {Karak}(2012)}]{CK12}
---. 2012, Physical Review Letters, 109, 171103

\bibitem[{{Covas} {et~al.}(1998){Covas}, {Tavakol}, {Tworkowski}, \&
  {Brandenburg}}]{CTTB98}
{Covas}, E., {Tavakol}, R., {Tworkowski}, A., \& {Brandenburg}, A. 1998, \aap,
  329, 350

\bibitem[{{Dasi-Espuig} {et~al.}(2010){Dasi-Espuig}, {Solanki}, {Krivova},
  {Cameron}, \& {Pe{\~n}uela}}]{Das10}
{Dasi-Espuig}, M., {Solanki}, S.~K., {Krivova}, N.~A., {Cameron}, R., \&
  {Pe{\~n}uela}, T. 2010, \aap, 518, A7

\bibitem[{{Fuchs} {et~al.}(1999){Fuchs}, {R{\"a}dler}, \& {Rheinhardt}}]{FRR99}
{Fuchs}, H., {R{\"a}dler}, K.-H., \& {Rheinhardt}, M. 1999, Astron. Nachr.,
  320, 129

\bibitem[{{Hartmann} \& {Noyes}(1987)}]{HN87}
{Hartmann}, L.~W., \& {Noyes}, R.~W. 1987, \araa, 25, 271

\bibitem[{{Haugen} {et~al.}(2004){Haugen}, {Brandenburg}, \& {Dobler}}]{Hau04}
{Haugen}, N.~E., {Brandenburg}, A., \& {Dobler}, W. 2004, \pre, 70, 016308

\bibitem[{{Hoyng}(1988)}]{H88}
{Hoyng}, P. 1988, \apj, 332, 857

\bibitem[{{Hoyng}(1993)}]{H93}
---. 1993, \aap, 272, 321

\bibitem[{{Hoyng} {et~al.}(1994){Hoyng}, {Schmitt}, \& {Teuben}}]{HST94}
{Hoyng}, P., {Schmitt}, D., \& {Teuben}, L.~J.~W. 1994, \aap, 289, 265

\bibitem[{{Hoyt} \& {Schatten}(1996)}]{HS96}
{Hoyt}, D.~V., \& {Schatten}, K.~H. 1996, \solphys, 165, 181

\bibitem[{{Hubbard} {et~al.}(2011){Hubbard}, {Rheinhardt}, \&
  {Brandenburg}}]{HRB11}
{Hubbard}, A., {Rheinhardt}, M., \& {Brandenburg}, A. 2011, \aap, 535, A48

\bibitem[{{K{\"a}pyl{\"a}} \& {Brandenburg}(2009)}]{KB09}
{K{\"a}pyl{\"a}}, P.~J., \& {Brandenburg}, A. 2009, \apj, 699, 1059

\bibitem[{{Karak}(2010)}]{Kar10}
{Karak}, B.~B. 2010, \apj, 724, 1021

\bibitem[{{Karak} \& {Choudhuri}(2011)}]{KC11}
{Karak}, B.~B., \& {Choudhuri}, A.~R. 2011, \mnras, 410, 1503

\bibitem[{{Karak} \& {Choudhuri}(2013)}]{KC13}
---. 2013, Res. Astron. Astrophys., 13, 1339

\bibitem[{{Karak} {et~al.}(2015){Karak}, {K{\"a}pyl{\"a}}, {K{\"a}pyl{\"a}},
  {Brandenburg}, {Olspert}, \& {Pelt}}]{Kar15}
{Karak}, B.~B., {K{\"a}pyl{\"a}}, P.~J., {K{\"a}pyl{\"a}}, M.~J.,
  {Brandenburg}, A., {Olspert}, N., \& {Pelt}, J. 2015, \aap, 576, A26

\bibitem[{{Kitchatinov} \& {Olemskoy}(2010)}]{KO10}
{Kitchatinov}, L.~L., \& {Olemskoy}, S.~V. 2010, Astron. Lett., 36, 292

\bibitem[{{Kitchatinov} {et~al.}(1994){Kitchatinov}, {R{\"u}diger}, \&
  {K{\"u}ker}}]{Kit94}
{Kitchatinov}, L.~L., {R{\"u}diger}, G., \& {K{\"u}ker}, M. 1994, \aap, 292,
  125

\bibitem[{{Kraft}(1967)}]{K67}
{Kraft}, R.~P. 1967, \apj, 150, 551

\bibitem[{{Krause} \& {R{\"a}dler}(1980)}]{KR80}
{Krause}, F., \& {R{\"a}dler}, K.~H. 1980, {Mean-field magneto\-hydro\-dynamics
  and dynamo theory} (Oxford: Pergamon Press)

\bibitem[{{K{\"u}ker} {et~al.}(1999){K{\"u}ker}, {Arlt}, \&
  {R{\"u}diger}}]{Kuk99}
{K{\"u}ker}, M., {Arlt}, R., \& {R{\"u}diger}, G. 1999, \aap, 343, 977

\bibitem[{{Moss} {et~al.}(1992){Moss}, {Brandenburg}, {Tavakol}, \&
  {Tuominen}}]{MBTT92}
{Moss}, D., {Brandenburg}, A., {Tavakol}, R., \& {Tuominen}, I. 1992, \aap,
  265, 843

\bibitem[{{Moss} {et~al.}(2008){Moss}, {Sokoloff}, {Usoskin}, \&
  {Tutubalin}}]{Mea08}
{Moss}, D., {Sokoloff}, D., {Usoskin}, I., \& {Tutubalin}, V. 2008, \solphys,
  250, 221

\bibitem[{{Nelson} {et~al.}(2013){Nelson}, {Brown}, {Brun}, {Miesch}, \&
  {Toomre}}]{Nel13}
{Nelson}, N.~J., {Brown}, B.~P., {Brun}, A.~S., {Miesch}, M.~S., \& {Toomre},
  J. 2013, \apj, 762, 73

\bibitem[{{Olemskoy} \& {Kitchatinov}(2013)}]{OK13}
{Olemskoy}, S.~V., \& {Kitchatinov}, L.~L. 2013, \apj, 777, 71

\bibitem[{{Ossendrijver} {et~al.}(1996){Ossendrijver}, {Hoyng}, \&
  {Schmitt}}]{OHS96}
{Ossendrijver}, A.~J.~H., {Hoyng}, P., \& {Schmitt}, D. 1996, \aap, 313, 938

\bibitem[{{Passos} {et~al.}(2014){Passos}, {Nandy}, {Hazra}, \&
  {Lopes}}]{Pas14}
{Passos}, D., {Nandy}, D., {Hazra}, S., \& {Lopes}, I. 2014, \aap, 563, A18

\bibitem[{{Racine} {et~al.}(2011){Racine}, {Charbonneau}, {Ghizaru}, {Bouchat},
  \& {Smolarkiewicz}}]{RCGBS11}
{Racine}, {\'E}., {Charbonneau}, P., {Ghizaru}, M., {Bouchat}, A., \&
  {Smolarkiewicz}, P.~K. 2011, \apj, 735, 46

\bibitem[{{Rempel} {et~al.}(2009){Rempel}, {Proctor}, \& {Chian}}]{Rea09}
{Rempel}, E.~L., {Proctor}, M.~R.~E., \& {Chian}, A.~C.-L. 2009, \mnras, 400,
  509

\bibitem[{{Rengarajan}(1984)}]{R84}
{Rengarajan}, T.~N. 1984, \apjl, 283, L63

\bibitem[{{R{\"u}diger} {et~al.}(1994){R{\"u}diger}, {Kitchatinov},
  {K{\"u}ker}, \& {Schultz}}]{Rea94}
{R{\"u}diger}, G., {Kitchatinov}, L.~L., {K{\"u}ker}, M., \& {Schultz}, M.
  1994, Geophys. Astrophys. Fluid Dynam., 78, 247

\bibitem[{{Ruzmaikin}(1981)}]{Ruz81}
{Ruzmaikin}, A.~A. 1981, Comm. Astrophys., 9, 85

\bibitem[{{Saar} \& {Baliunas}(1992)}]{SB92}
{Saar}, S.~H., \& {Baliunas}, S.~L. 1992, in Astron. Soc. Pac. Conf. Ser.,
  Vol.~27, The Solar Cycle, ed. K.~L. {Harvey}, 150--167

\bibitem[{{Sokoloff} \& {Nesme-Ribes}(1994)}]{SN94}
{Sokoloff}, D., \& {Nesme-Ribes}, E. 1994, \aap, 288, 293

\bibitem[{{Spiegel}(1977)}]{Spi77}
{Spiegel}, E.~A. 1977, in Lecture Notes in Physics, Berlin Springer Verlag,
  Vol.~71, Problems of Stellar Convection, ed. E.~A. {Spiegel} \& J.-P. {Zahn},
  267--283

\bibitem[{{Tavakol}(1978)}]{Tav78}
{Tavakol}, R.~K. 1978, \nat, 276, 802

\bibitem[{{Usoskin}(2013)}]{Uso13}
{Usoskin}, I.~G. 2013, Living Reviews in Solar Physics, 10, 1

\bibitem[{{Usoskin} {et~al.}(2007){Usoskin}, {Solanki}, \& {Kovaltsov}}]{USK07}
{Usoskin}, I.~G., {Solanki}, S.~K., \& {Kovaltsov}, G.~A. 2007, \aap, 471, 301

\bibitem[{{Usoskin} {et~al.}(2014){Usoskin}, {Hulot}, {Gallet}, {Roth},
  {Licht}, {Joos}, {Kovaltsov}, {Th{\'e}bault}, \& {Khokhlov}}]{Uea14}
{Usoskin}, I.~G., {et~al.} 2014, \aap, 562, L10

\bibitem[{{Weiss} {et~al.}(1984){Weiss}, {Cattaneo}, \& {Jones}}]{WCJ84}
{Weiss}, N.~O., {Cattaneo}, F., \& {Jones}, C.~A. 1984, Geophys. Astrophys.
  Fluid Dynam., 30, 305

\end{thebibliography}

\end{document}